\documentclass[aps,prl,showpacs,superscriptaddress,amsmath,amssymb,twocolumn,preprintnumbers,reprint]{revtex4}
\usepackage{graphicx}
\usepackage{dcolumn}
\usepackage{bm}
\usepackage{amsmath}
\usepackage{amssymb}
\usepackage[english]{babel}
\usepackage{hyperref}
\usepackage{natbib}
\usepackage{xcolor}
\hypersetup{colorlinks=true,linkbordercolor=blue,linkcolor=blue, citecolor=blue,pdfborderstyle={/S/U/W 1}}
\usepackage{url}
\def\url#1{\href{#1}{{\tt #1}}}

\def\be{\begin{equation}}
 \def\ee{\end{equation}}
\def\X{{\rm X}}
\newcommand{\mrm}[1]{{\mathrm{#1}}}
  \newcommand{\eg}{{\em e.g.}}
\newcommand{\eq}[1]{(\ref{#1})}
\newcommand{\jpsi}{J/\psi}
\newcommand{\xf}{x_{\mathrm{F}}}
\newcommand{\xtwo}{x_{2}}
\newcommand{\pp}{p--p}
\newcommand{\pA}{p--A}
\newcommand{\dd}{{\rm d}}
\newcommand{\lsim}{\lesssim} 
\newcommand{\qzero}{\hat{q}_0}
\newcommand{\gevsqfm}{GeV$^2$/fm}
\newcommand{\A}{{\rm A}}
\newcommand{\sqrts}{\sqrt{s}}
\newcommand{\pPb}{p--Pb}
\newcommand{\Pb}{{\rm Pb}}
\newcommand{\gev}{{\mrm{GeV}}}

\begin{document}
\title{Disentangling Shadowing from Coherent Energy Loss using the Drell-Yan Process}
\author{Fran\c{c}ois Arleo}
\affiliation{Laboratoire Leprince-Ringuet, \'Ecole polytechnique, CNRS/IN2P3,  Universit\'e Paris-Saclay,  91128, Palaiseau, France}
\author{St\'ephane Peign\'e}
\affiliation{SUBATECH, 
Universit\'e de Nantes, Ecole des Mines de Nantes, CNRS/IN2P3, 4 rue Alfred Kastler, 44307 Nantes cedex 3, France}

\date{\today}

\begin{abstract}
We suggest the measurement of Drell-Yan (DY) lepton pairs in \pPb\ collisions at the LHC ($\sqrts=5.02$~TeV) in order to disentangle the relative contributions of shadowing and coherent energy loss in quarkonium production off nuclei. The nuclear modification of low mass DY production is computed at NLO using various sets of nuclear parton densities. It is then observed that shadowing effects strongly cancel out in the $\jpsi$ over DY suppression ratio
$R_{\rm pA}^{\psi}(y)\big/R_{\rm pA}^{\rm DY}(y)$, unlike the effect of coherent energy loss. Such a measurement could be performed at forward rapidity by the ALICE and LHCb collaborations at the LHC.
\end{abstract}
\maketitle

\setcounter{footnote}{0}
\renewcommand{\thefootnote}{\arabic{footnote}}

Measurements of $\jpsi$ production in \pPb\ collisions at the LHC ($\sqrts=5.02$~TeV) by ALICE~\cite{Abelev:2013yxa,Adam:2015iga} and LHCb~\cite{Aaij:2013zxa}, and the observation of a strong attenuation at large rapidity with respect to the \pp\ data interpolation at the same collision energy, has triggered an intense debate on the origin of such a nuclear suppression~\cite{Andronic:2015wma}. Several groups have attributed the suppression to the depletion of the gluon distribution in the target nucleus expected at small $\xtwo \lesssim 10^{-2}$. Such a depletion, commonly named `shadowing', is currently either incorporated in nuclear parton distribution functions (nPDFs) obtained from global fits based on DGLAP evolution, or determined from non-linear QCD evolution within the saturation formalism (see~\cite{Armesto:2006ph,Gelis:2010nm} for topical reviews). However another fundamental phenomenon, namely coherent parton energy loss in cold nuclear matter~\cite{Arleo:2010rb,Arleo:2012rs,Armesto:2012qa,Arleo:2013zua,Armesto:2013fca,Peigne:2014uha,Liou:2014rha,Peigne:2014rka}, could explain these data. It is therefore crucial to quantify the relative effects of shadowing and coherent energy loss. In this Letter, we show that the measurement of Drell-Yan (DY) lepton pairs of relatively low mass ($10\lesssim M_{\rm DY}\lesssim 20$~GeV) in \pA\ collisions could be decisive to reach this goal. As a consequence, this measurement would shed light on the production of hadrons in \pA\ collisions, expected to be sensitive to coherent energy loss~\cite{Arleo:2010rb} on top of possible shadowing effects.

Coherent energy loss and shadowing are two distinct effects, and should in principle be both taken into account in nuclear suppression models. However, as discussed in~\cite{Arleo:2012rs,Arleo:2013zua} coherent energy loss \emph{alone} allows one to describe $\jpsi$ nuclear suppression observed at large $\xf$ at fixed-target collision energies, $\sqrts \lsim 40$~GeV~\cite{Leitch:1999ea}, with a value of the cold nuclear matter transport coefficient $\qzero=0.07-0.09$~\gevsqfm. This is consistent with the fact that shadowing is expected to be small at those energies. Taking in addition into account the shadowing given by either EPS09~\cite{Eskola:2009uj} or DSSZ~\cite{deFlorian:2011fp} next-to-leading order (NLO) nPDF sets these data can be described with a slightly smaller value of $\qzero$, although energy loss remains the dominant effect at fixed-target energies~\cite{Arleo:2012rs}. Extrapolating coherent energy loss to collider energies leads to a central  prediction~\cite{Arleo:2012rs} (with a rather narrow theoretical uncertainty band~\cite{Andronic:2015wma}) which agrees well with the $\jpsi$ suppression data measured in d--Au collisions at RHIC~\cite{Adare:2010fn,Adare:2012qf} and \pPb\ collisions at the LHC~\cite{Abelev:2013yxa,Aaij:2013zxa}. Although experimental uncertainties still leave room for shadowing in $\jpsi$ suppression at collider energies, the results of the pure energy loss scenario tend to favor nPDF sets with a moderate shadowing.

Regarding calculations including only shadowing, a prediction at NLO within the Color Evaporation Model (CEM) for quarkonium production using EPS09 NLO is in reasonable agreement with the data, except in the most forward bins where the measured suppression is stronger than predicted~\cite{Albacete:2013ei} (these calculations were updated in~\cite{Vogt:2015uba}). After the data came, a leading-order (LO) Color Singlet Model (CSM) calculation~\cite{Ferreiro:2013pua} using EPS09 LO proved to be in apparent agreement with data; however in this case the theoretical uncertainty associated with the use of EPS09 LO appears very large. The only genuine prediction in the saturation formalism~\cite{Fujii:2013gxa} turns out to largely overestimate $\jpsi$ nuclear suppression. Using the same model but with an improved treatment of the non-linear QCD evolution, a more recent calculation~\cite{Ducloue:2015gfa} led to less $\jpsi$ suppression at forward rapidity, hence to lesser disagreement with experimental results. Finally, saturation effects have also been investigated within the Non-Relativistic QCD (NRQCD) formalism, in which the resulting theoretical uncertainty band is large and encompasses the data~\cite{Ma:2015sia}. Clearly, the large theoretical uncertainties of nPDF and saturation calculations do not yet allow for a clear interpretation of the $\jpsi$ measurements at the LHC.

Drell-Yan forward production in \pPb\ collisions at the LHC offers a unique opportunity to clarify this si\-tua\-tion. Shadowing effects on both $\jpsi$ and DY are expected to be of similar ma\-gni\-tude, since  small-$\xtwo$ gluons and sea antiquarks in nuclei --~contributing respectively to $\jpsi$ and DY production at large enough rapidity~-- have a similar depletion in most nPDF global fits, $R_g^{\A} \simeq R_{\bar{q}}^{\A}$ (with $R_i^{\A}(x, Q)\equiv f_i^{\rm p/A}(x, Q)/f_i^{\rm p}(x, Q)$, where  $f^{\rm p}$ and $f^{\rm p/A}$ are respectively the PDF in a free proton and in a proton bound in nucleus A). Consequently, comparing the $\jpsi$ and DY nuclear modification factor in \pA\ collisions,
\be
\label{eq:rpa}
R_{\rm pA}(y) \equiv \frac{1}{A}\ \frac{\dd\sigma_{\rm pA}}{\dd{y}}\ \Big/\ \frac{\dd\sigma_{\rm pp}}{\dd{y}}\ ,
\ee
for instance through the measurement of the double ratio, 
\be
\label{double-ratio}
{\cal R}_{\rm pA}^{\psi/{\rm DY}}(y) \equiv R_{\rm pA}^{\psi}(y)\ \big/\ R_{\rm {pA}}^{\rm DY}(y)\ ,
\ee
allows for an important ``cancellation'' of nPDF effects (and most likely, saturation effects too). Remarkably, when it comes to coherent energy loss effects, no such cancellation is expected.

Indeed, coherent energy loss arises from the interference between initial and final state gluon radiation and therefore affects, in \pA\ collisions, only those partonic subprocesses with a {\it colorful} final state~\cite{Peigne:2014uha}. Coherent energy loss is thus present in $\jpsi$ production, where the $c \bar{c}$ pair is produced either as a color octet or as part of a {\it compact} octet system (like $c\bar{c} + g$ in the CSM), but absent in the DY subprocess at leading order, $q \bar{q} \to \gamma^\star$. At NLO, the virtual photon is produced together with an additional parton, mainly through $q g \to q\gamma^{\star}$ at large $y$, making DY production potentially sensitive to coherent energy loss. However, the medium-induced coherent radiation spectrum associated with $q g \to q\gamma^{\star}$ is small ($\propto 1/N_c$) and moreover {\it negative}~\cite{Peigne:2014uha}, leading to a slight DY {\it enhancement}, at variance with $\jpsi$ suppression.

In summary, the qualitative expectations for the $\jpsi$ over DY suppression at large rapidity are as follows:
\begin{center} 
\vspace{-3mm}
\begin{tabular}{p{0.11\textwidth}p{0.18\textwidth}p{0.025\textwidth}p{0.17\textwidth}} 
\hskip 5mm {\rm \bf nPDF} & $R^{\psi} \simeq {R}^{{\rm DY}}$ &$\rightarrow$ & ${\cal R}^{\psi/{\rm DY}}\simeq 1$  \\
\hskip 5mm {\rm \bf  E.~loss} & $R^{\psi} < 1$; ${R}^{{\rm DY}} \gtrsim 1$ &$\rightarrow$ & ${\cal R}^{\psi/{\rm DY}}< 1$
\end{tabular} 
\end{center} 
This is supported by the quantitative discussion below. 

%%%%%%%%%%%%%%%%%%%%%%%%%%%%%%%%%%%%%%%%%%
\begin{figure}[tbp]
\begin{center}
    \includegraphics[width=8.8cm]{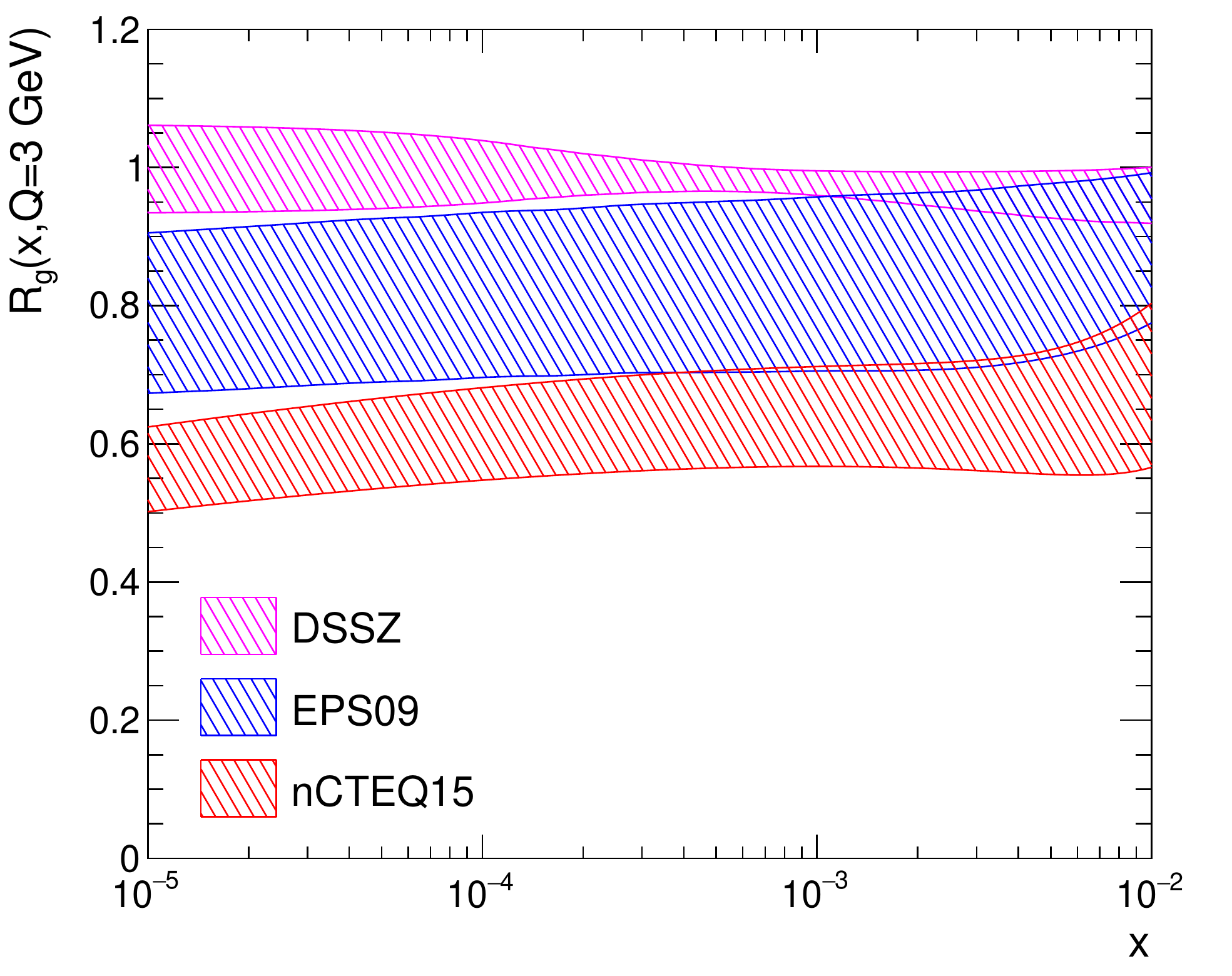}
    \includegraphics[width=8.8cm]{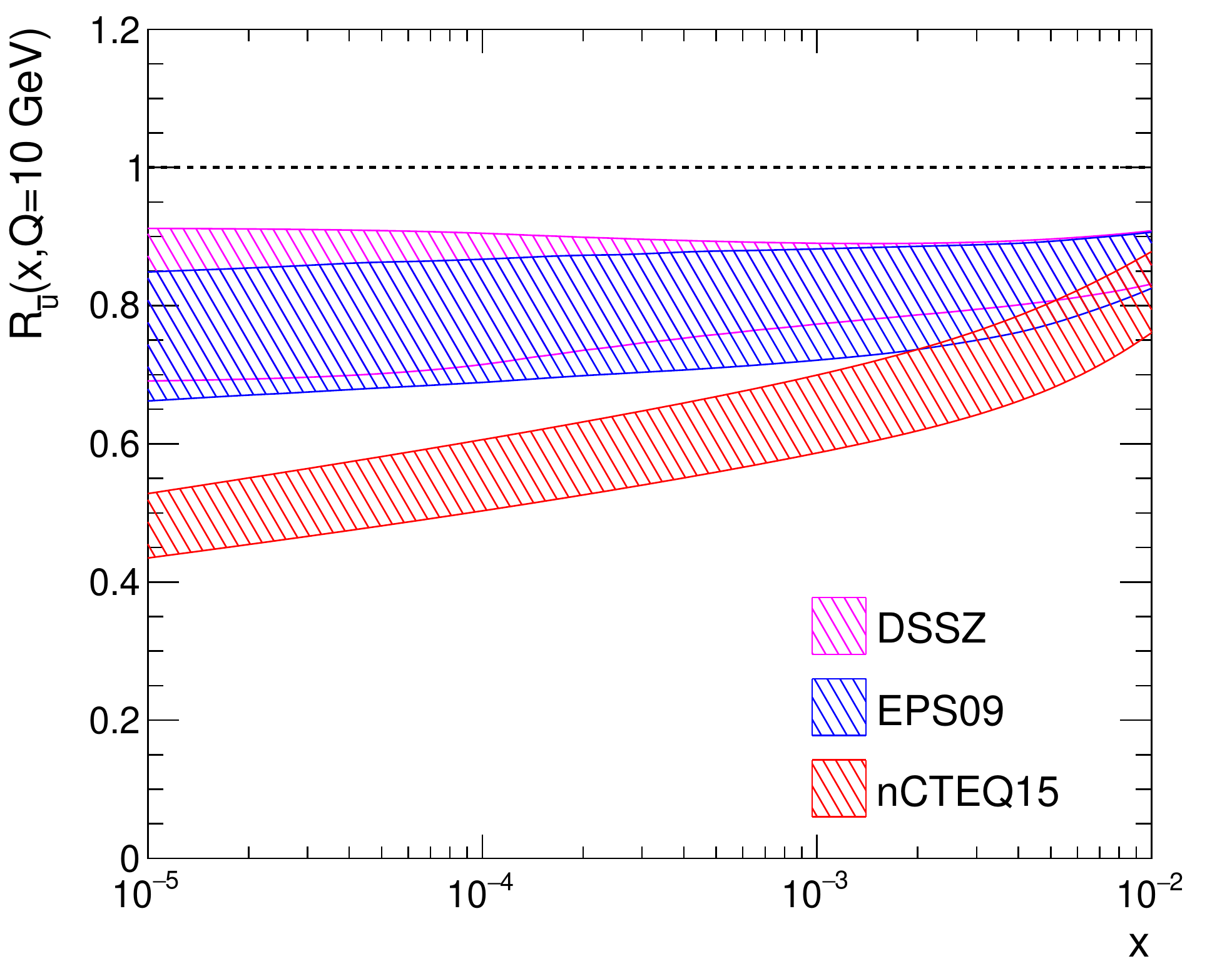}
  \end{center}
\caption{$R^{\Pb}_g(x, Q=3~\gev)$ (top) and $R^{\Pb}_{\bar{u}}(x, Q=10~\gev)$ (bottom) for various nPDF sets.}
   \label{fig:npdf}
\end{figure}
%%%%%%%%%%%%%%%%%%%%%%%%%%%%%%%%%%%%%%%%%%

Before moving to the actual results on DY and $\jpsi$ nuclear suppression, we first illustrate in Fig.~\ref{fig:npdf} the gluon (top) and $\bar{u}$-quark (bottom) nPDF ratios  (for a lead target) and their uncertainties, given by the three most recent NLO nPDF sets from global fits: EPS09~\cite{Eskola:2009uj}, DSSZ~\cite{deFlorian:2011fp}, nCTEQ15~\cite{Kovarik:2015cma}. (Results using the earlier HKN07 set~\cite{Hirai:2007sx} are very similar to those obtained with the central set of EPS09 and are therefore not reproduced here.) The choice $Q=3~{\rm GeV}$ (resp.~$Q=10~{\rm GeV}$) for the factorization scale in $R_{g}$ (resp.~$R_{\bar{u}}$) and the small-$x$ range, $10^{-5} < x < 10^{-2}$, reflect our proposal to compare $\jpsi$ to DY pairs of mass above that of the $\Upsilon$ states, in the rapidity range $0 < y < 5$ in \pPb\ collisions at the LHC. The bands are determined from the spread of 30, 50, and 32 uncertainty sets around the central prediction of EPS09, DSSZ and nCTEQ15, respectively. The current uncertainty on the small-$x$ gluon shadowing is striking (Fig.~\ref{fig:npdf}, top). Depending on which set is used, the nPDF ratio at $x=10^{-5}$ varies from $R_g \simeq 0.95$--$1.05$ in DSSZ to $R_g \simeq 0.5$--$0.6$ in nCTEQ15. Moreover the uncertainty bands are rather broad and almost do not overlap in the entire $x$-domain, reflecting the fact that these should be seen as \emph{lower} estimates for the theoretical uncertainty, as discussed in~\cite{Eskola:2009uj,Kovarik:2015cma}. We also note that for EPS09 and nCTEQ15, the shadowing of sea antiquarks is of the same magnitude as that of gluons. Although not directly apparent in the uncertainty bands of Fig.~\ref{fig:npdf}, the latter statement holds separately for \emph{each} uncertainty set of EPS09 and nCTEQ15. As a consequence, the double ratio $R_g/R_{\bar{u}}$ is close to unity, with an uncertainty band much smaller than that of the single nPDF ratios $R_g$ and $R_{\bar{u}}$. This will translate into rather precise nPDF predictions for the double ratio ${\cal R}^{\psi/{\rm DY}}$ using EPS09 and nCTEQ15 sets. In the particular case of DSSZ, gluon and sea antiquark shadowing do not have similar magnitudes ($R_g\simeq0.95$--$1.05$ whereas $R_{\bar{u}}\simeq 0.7$--$0.9$) and the uncertainty band for ${\cal R}^{\psi/{\rm DY}}$ will remain large but above unity, hence even further away from the prediction of the coherent energy loss model.

\begin{figure}[t]
\begin{center}
    \includegraphics[width=8.8cm]{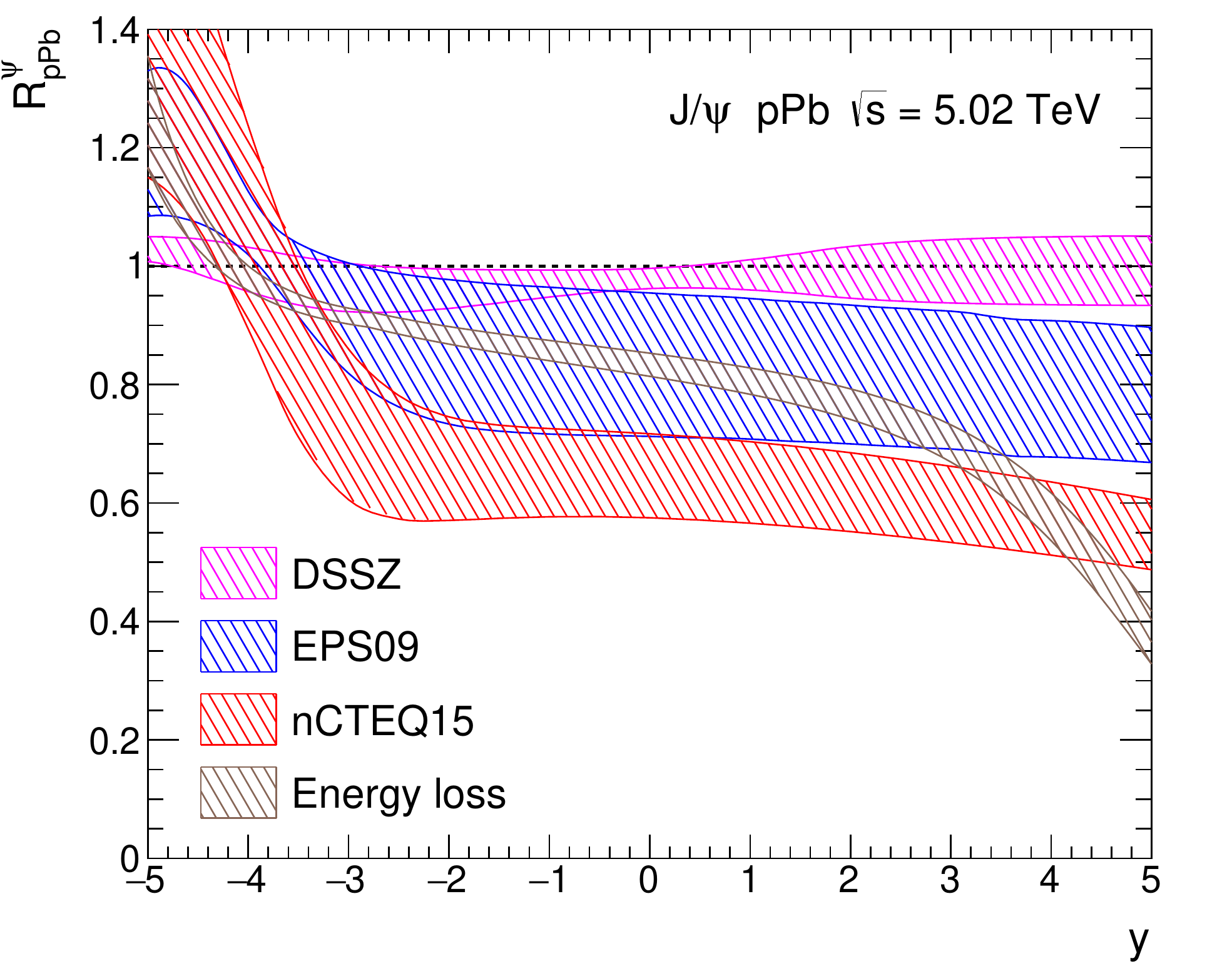}
    \includegraphics[width=8.8cm]{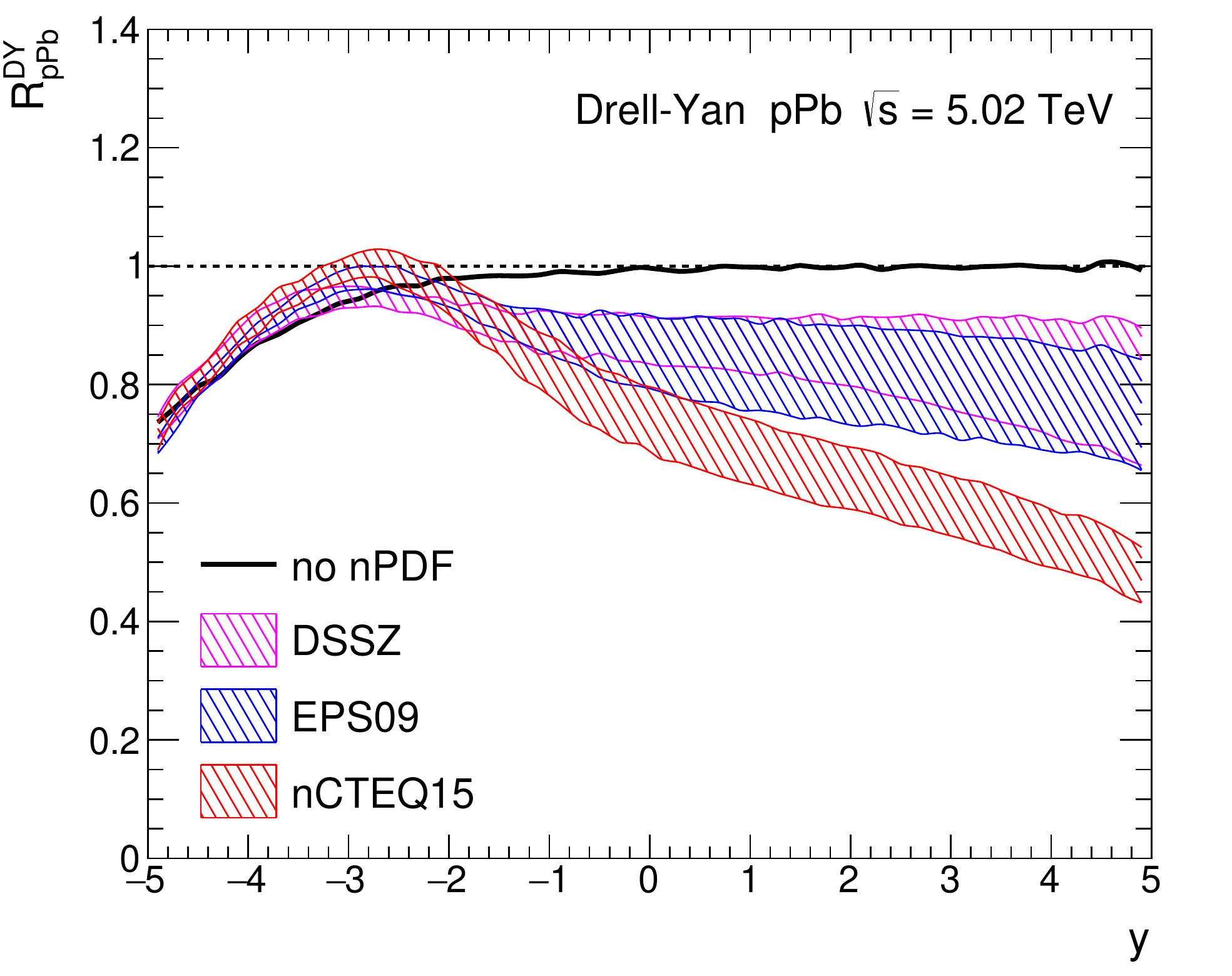}
  \end{center}
\caption{$\jpsi$ (top) and DY (bottom) suppression in \pPb\ collisions at $\sqrts=5.02$~TeV for the various nPDF sets used in this study. 
The prediction for $\jpsi$ suppression from the effect of coherent energy loss alone is also shown (top).}
  \label{fig:psi_dy}
\end{figure}

Let us now discuss shadowing effects on $\jpsi$ suppression. Unlike DY production, for which perturbative calculations are well established, quarkonium production in hadronic collisions is not yet fully understood, making the calculation of {\it absolute} production cross sections uncertain. However, the magnitude of shadowing effects on the $\jpsi$ nuclear modification factor \eq{eq:rpa} should be under better control. Indeed, in most approaches (CEM, CSM, NRQCD), $\jpsi$ production at the LHC is dominated by gluon fusion, $gg\to \jpsi+\X$. Therefore shadowing effects should be dictated by the {\it gluon} nPDF ratio, $R_g^{\rm Pb}(\xtwo, Q)$, quite independently of the production mechanism. The specific  $\jpsi$ production mechanism may nevertheless affect the kinematics of the process and thus the typical value of $\xtwo$. However, as can be seen in Fig.~\ref{fig:npdf} (top), $R_g^{\rm Pb}$ is extremely flat at small values of $x$ (note the logarithmic scale), for all nPDF sets; the uncertainties associated with the gluon distributions in nuclei exceed by far those arising from the uncertainty on $\xtwo$. For the sake of simplicity, we shall thus assume that $\jpsi$ suppression due to shadowing is given by $R_{\rm pPb}^{\psi}=R_g^{\rm Pb}(\xtwo, Q=M_{\psi})$, where $\xtwo$ is given by the LO expression $x_2 = {M_{\psi}} \, e^{-y}/{\sqrts}$. Fig.~\ref{fig:psi_dy} (top) displays shadowing effects on $\jpsi$ suppression in \pPb\ collisions at the LHC, exhibiting the large uncertainty at forward rapidity. The predictions for $\jpsi$ suppression from coherent energy loss {\it alone} are also shown in Fig.~\ref{fig:psi_dy} (top), which illustrates the difficulty to trace the physical origin of $\jpsi$ suppression seen in the LHC data~\footnote{Although the energy loss model predictions shown in Fig.~\ref{fig:psi_dy} (top) do not include shadowing effects, note however that the transport coefficient is varied in a rather large interval, $\qzero=0.05$--$0.09$~\gevsqfm, in order to encompass the smaller values of $\qzero$  extracted at fixed target energies when attributing part of the $\jpsi$ suppression to the (small) amount of shadowing expected at those energies~\cite{Arleo:2012rs}.}.  Finally, let us remark that the 
uncertainty band of Fig.~\ref{fig:psi_dy} (top) obtained for EPS09 is very similar to that found in \cite{Vogt:2015uba} (see Fig.~8 therein) using explicit calculations within the CEM at NLO.

%%%%%%%%%%%%%%%%%%%%%%%%%%%%%%%%%%%%%%%%%%
%%%%%%%%%%%%%%%%%%%%%%%%%%%%%%%%%%%%%%%%%%

We now come the main point of this Letter, namely, how the study of the DY nuclear modification factor can help clarify this situation. In the following, we first determine the nPDF effects on the DY nuclear modification factor defined in \eq{eq:rpa}, and then discuss the double ratio \eq{double-ratio}. Calculations are done at NLO using the DYNNLO~\cite{Catani:2007vq,Catani:2009sm} Monte Carlo program. The single differential cross section ${\dd}\sigma/{\dd}y$ is computed in \pp\ and \pPb\ collisions, from which $R_{\rm {pPb}}^{\rm DY}(y)$ is determined. We use the MSTW NLO~\cite{Martin:2009iq} proton PDF, and factorization and renormalization scales equal to the DY mass $M_{\rm DY}$. The \pPb\ calculations were carried out using the NLO nPDF sets already discussed. For completeness, the DY cross section has also been computed in \pPb\ collisions assuming no nPDF corrections ($R^{\rm Pb}_i(x, Q)=1$). The mass range considered in this calculation, $10.5 < M_{\rm DY} < 20$~GeV, appears as an interesting compromise. On the one hand, $M_{\rm DY}$ should not be too large, both to guarantee a good cancellation of nPDF effects between $\jpsi$ and DY suppression, and moreover to ensure a reasonable statistics at the LHC. On the other hand, for $M_{\rm DY} < 10.5$~GeV the extraction of the DY signal is extremely delicate, due to the large background of lepton pairs coming from heavy-flavor hadron decays (adding to those from quarkonia up to the $\Upsilon$(3S) of mass $10.35$~GeV)~\cite{LHCb:2012fja}. Note that this mass range has also been considered in saturation studies~\cite{GolecBiernat:2010de,Ducati:2013cga} although no DY nuclear modification factor was determined.

The DY suppression in \pPb\ collisions is shown in Fig.~\ref{fig:psi_dy} (bottom) as a function of the lepton pair rapidity. In the most forward bins, $3 \lesssim y \lesssim 5$ (corresponding to $10^{-5} \lesssim x_2 \lesssim 10^{-4}$ using $x_2 = {M_{\rm DY}}\,e^{-y}/{\sqrts}$), the similarity with the sea antiquark nPDF ratios shown in Fig.~\ref{fig:npdf} (bottom) is clear: DY suppression is quite strong ($R_{\rm pPb}\simeq 0.5$--$0.6$) using nCTEQ15, and less pronounced using DSSZ or EPS09 ($R_{\rm pPb}^{\rm DY}\simeq 0.7$--$0.9$). Since no coherent energy loss (but a slight energy {\it gain}, as mentioned previously) is expected in DY production, these calculations already demonstrate the discriminating power of low-mass DY pair production in \pPb\ collisions at the LHC, allowing for setting tight constraints on antiquark shadowing at very small $x$. Such measurement could be performed by either ALICE or LHCb (as demonstrated by the early results in \pp\ collisions~\cite{LHCb:2012fja}), which dimuon rapidity acceptance extends up to $y_{\rm lab}=4.5$ (this corresponds to $y=y_{\rm lab}-\Delta y\simeq4$, where $\Delta y \simeq 0.465$ is the boost of the lab frame with respect to the center-of-mass frame). Moreover, counting rates are expected to be large. The \pPb\ cross section is ${\dd}\sigma^{\rm DY}_{\rm pPb}/{\dd}y \simeq 40$~nb in the rapidity bin $3.5<y<4$. Using an integrated luminosity of ${\cal L}^{\rm int}_{\rm pPb}=100$~nb$^{-1}$ at LHC~Run~2 (typically a few times larger than at Run~1), approximately ${\cal N}_{3.5<y<4}=2000$ pairs are expected to be produced in that rapidity bin. This ensures the statistical uncertainties on the ratio $R_{\rm pPb}^{\rm DY}$ to remain under control, at a few percent level, even at large rapidity. The backward region ($y<0$), where the depletion of DY production in \pPb\ with respect to \pp\ collisions is due to isospin effects~\footnote{At large negative rapidity the neutrons present in the lead nucleus are less effective than protons in producing DY pairs, since $\bar{u} u \to\gamma^\star$ dominates over $\bar{d}d\to\gamma^\star$ (because of the respective electric charges, $e_u^2/e_d^2=4$). At positive rapidity, DY production is dominated by sea (anti)quarks in the nucleus, and isospin effects are negligible. Isospin effects were discussed previously in the case of photon production in \pA\ collisions, see \eg~\cite{Arleo:2006xb}.}, would also be interesting in itself.

\begin{figure}[t]
\begin{center}
    \includegraphics[width=8.8cm]{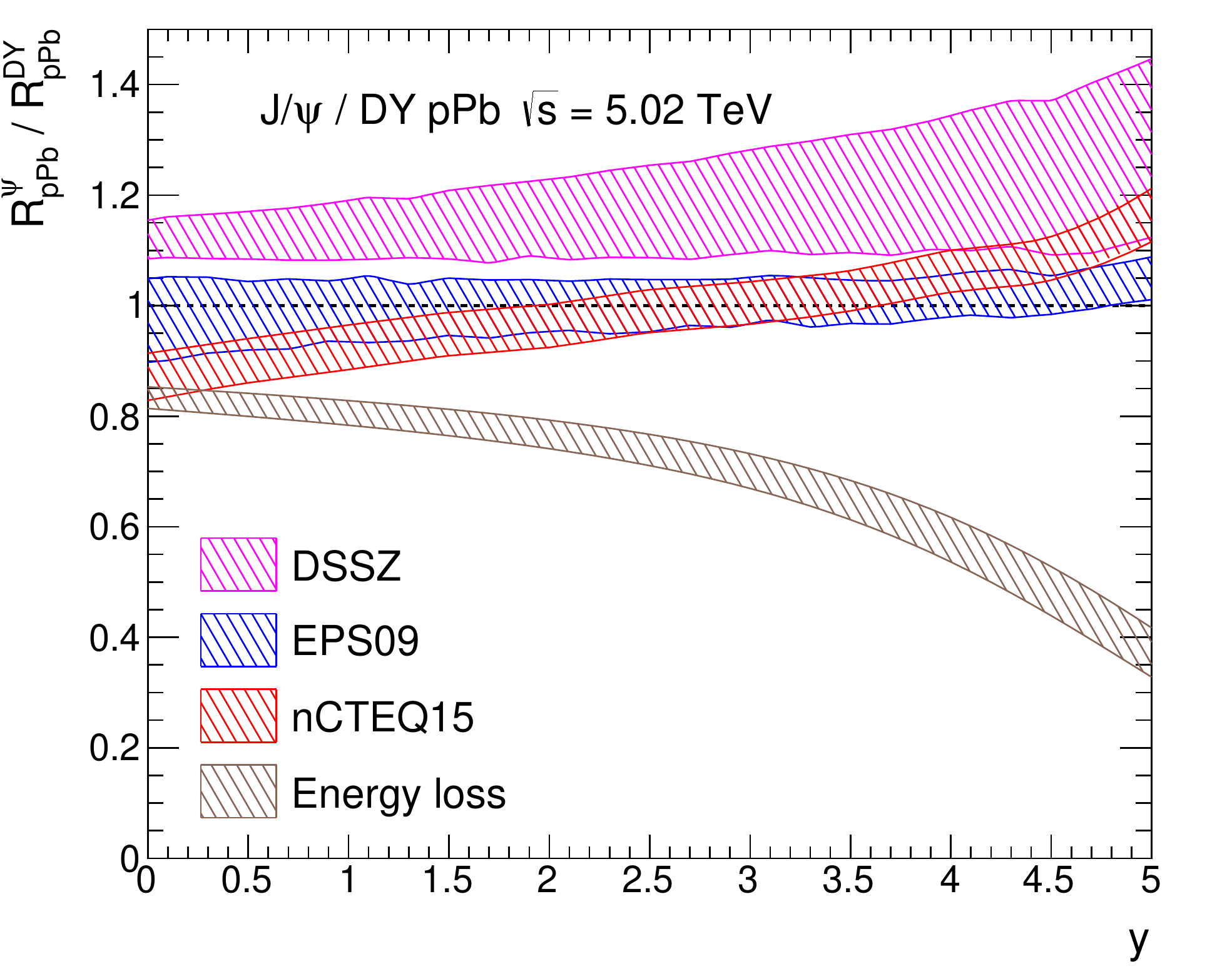}
  \end{center}
\caption{Double ratio ${\cal R}_{\rm pPb}^{\psi/{\rm DY}}$ in \pPb\ collisions at $\sqrts=5.02$~TeV for the various nPDF sets and in the coherent energy loss model.}
  \label{fig:psi_dy_ratio}
\end{figure}

Finally, in the forward rapidity domain the DY nuclear suppression factor, Fig.~\ref{fig:psi_dy} (bottom), compared to that of $\jpsi$, Fig.~\ref{fig:psi_dy} (top), can be used to disentangle coherent energy loss from shadowing effects, taking advantage of the similarity between gluon and sea antiquark shadowing as well as of the absence (or smallness) of coherent energy loss effects in DY production. Shadowing effects on the {\it double ratio} ${\cal R}_{\rm pA}^{\psi/{\rm DY}}$ defined in \eq{double-ratio} are shown in Fig.~\ref{fig:psi_dy_ratio}. The contrast with the {single ratio} $R_{\rm pA}^{\psi}$ is striking. Indeed, quite independently of the nPDF set, shadowing leads to a double ratio close to unity at forward rapidity (and slightly above in DSSZ, ${\cal R}_{\rm pA}^{\psi/{\rm DY}}\simeq1.1$--$1.3$ at $y=4$). Moreover, there is a rather small associated uncertainty, resulting from using the {\it same} nPDF uncertainty set for both $\jpsi$ and DY production. Assuming no coherent energy loss in DY production, the predictions of the sole effect of coherent energy loss on the double ratio are also shown in Fig.~\ref{fig:psi_dy_ratio}, and differ significantly from the calculations based on shadowing effects only. For the illustration one expects, at $y=4$, ${\cal R}_{^{\rm pA}}^{_{\psi/{\rm DY}}}\lesssim 0.6$ for coherent energy loss effects while ${\cal R}_{^{\rm pA}}^{_{\psi/{\rm DY}}}\simeq 1$--$1.3$ for nPDF effects. Comparing Fig.~\ref{fig:psi_dy_ratio} to Fig.~\ref{fig:psi_dy} (top) emphasizes the discriminating power of the double ratio ${\cal R}_{^{\rm pA}}^{_{\psi/{\rm DY}}}$ in \pA\ collisions at the LHC. Finally we stress that the prediction for the double ratio arising from the {\it combined} effects of coherent energy loss and shadowing, is roughly given by the product of the two~\cite{Arleo:2012rs}. It is thus either almost identical to the prediction assuming coherent energy loss only (when using EPS09 or nCTEQ15) or at most enhanced by 20\%--30\% (when using DSSZ).

%%%%%%%%%%%%%%%%%%%%%%%%%%%%%%%%%%%%%%%%%%
%%%%%%%%%%%%%%%%%%%%%%%%%%%%%%%%%%%%%%%%%%

In summary, we propose to use the Drell-Yan process and in particular the double ratio ${\cal R}_{\rm pA}^{\psi/{\rm DY}}$ to disentangle the two main effects expected to be at work in $\jpsi$ production (and more generally in hadron production) in nuclear collisions. Such a measurement, which could be performed by ALICE and LHCb, would shed light on the physical origin of the current data, therefore entailing significant consequences for the interpretation of $\jpsi$ suppression also in Pb--Pb collisions.

\begin{acknowledgements}
This work is funded by ``Agence Nationale de la Recherche'' under grant ANR-PARTONPROP.
\end{acknowledgements}

\end{document}